\documentclass[useAMS,usenatbib,times,a4paper,amssymb]{mn2e}
\usepackage{epsfig,times,amssymb,amsmath}

\title[Atmospheric Distortions]{The impact of high spatial frequency atmospheric distortions on weak lensing measurements}
\author[C. Heymans \& B. Rowe et al.]{Catherine Heymans,$^{1}$\thanks{email: heymans@roe.ac.uk \& browe@star.ucl.ac.uk} Barnaby Rowe,$^{2,3}$$^{\star}$ Henk Hoekstra,$^{4}$ Lance Miller,$^5$ \newauthor
Thomas Erben,$^6$ Thomas Kitching,$^1$ \& Ludovic Van Waerbeke$^7$
\\
$^1$Scottish Universities Physics Alliance, Institute for Astronomy, University of Edinburgh, Royal Observatory, Blackford Hill, Edinburgh, EH9 3HJ, UK. \\ 
$^2$Department of Physics \& Astronomy, University College London, Gower Street, London, London, WC1E 6BT, UK.\\
$^3$California Institute of Technology, 1200 E California Boulevard, Pasadena, CA, 91106, USA.\\
$^{4}$Leiden Observatory, Leiden University, Niels Bohrweg2, NL-2333 CA Leiden, The Netherlands. \\  
$^5$Department of Physics, Oxford University, Denys Wilkinson Building, Keble Road, Oxford OX1 3RH, UK. \\
$^{6}$Argelander-Institut f\"ur Astronomie, Universit\"at  Bonn, Auf dem H\"ugel 71, 53121 Bonn, Germany.\\
$^7$University of British Columbia, 6224 Agricultural Rd., Vancouver, BC, V6T 1Z1, Canada.\\  
}

\newcommand{\be}{\begin{equation}}  \newcommand{\ee}{\end{equation}}
  \newcommand{\ba}{\begin{eqnarray}} \newcommand{\ea}{\end{eqnarray}}
\newcommand{\bm}[1]{\mbox{\boldmath{$#1$}}}

\begin{document}

\maketitle

\begin{abstract}
High precision cosmology with weak gravitational lensing requires a precise measure of the Point Spread Function across the imaging data where the accuracy to which high spatial frequency variation can be modeled is limited by the stellar number density across the field.  We analyse dense stellar fields imaged at the Canada-France-Hawaii Telescope to quantify the degree of high spatial frequency variation in ground-based imaging Point Spread Functions and compare our results to models of atmospheric turbulence.  The data shows an anisotropic turbulence pattern with an orientation independent of the wind direction and wind speed.  We find the amplitude of the high spatial frequencies to decrease with increasing exposure time as $t^{-1/2}$, and find a negligibly small atmospheric contribution to the Point Spread Function ellipticity variation for exposure times $t>180$ seconds.  For future surveys analysing shorter exposure data, this anisotropic turbulence will need to be taken into account as the amplitude of the correlated atmospheric distortions becomes comparable to a cosmological lensing signal on scales less than $\sim 10$ arcminutes.  This effect could be mitigated, however, by correlating galaxy shear measured on exposures imaged with a time separation greater than 50 seconds, for which we find the spatial turbulence patterns to be uncorrelated.  
\end{abstract}

\begin{keywords}
cosmology: observations - gravitational lensing 
\end{keywords}

\section{Introduction}
{Over the next decade we will see all sky surveys being undertaken to address key science questions ranging from cataloging all near earth objects to mapping the Dark Universe.   These surveys include the Large Synoptic Survey Telescope \citep[LSST]{LSST09}, Euclid \citep{Euclid10}, and the Panoramic Survey Telescope and Rapid Response System \citep[PanSTARRS]{PS10}.   By observing the growth of dark matter structures over cosmic time, weak lensing may provide the most interesting constraints on the dynamical properties of dark energy.  Alongside modifications to General Relativity, this dark energy is hypothesized as an explanation for the accelerated expansion of the Universe \citep[see for example][and references therein]{Huterer10}.  The observational measurement of weak gravitational lensing is however far from trivial.  In fact it is the technique that sets the most stringent requirements on the optical design of future instrumentation as we require accurate knowledge of the Point Spread Function (PSF).  The PSF also needs to be minimal in terms of its size and ellipticity.}  Dark matter typically induces a one percent change in the ellipticity of lensed distant galaxies. It is this distortion that we measure to constrain cosmology.  In contrast the telescope, camera and atmosphere can introduce image distortions up to the ten percent level.  The main challenge in weak lensing is therefore to distinguish the cosmological distortion from instrumental and atmospheric distortions in order to recover the underlying dark matter signal.  A significant effort is underway to improve the accuracy of weak lensing measurements both from a software \citep{GREAT10} and hardware perspective \citep{LSST09}.  Furthermore ambitious space-based surveys are proposed in order to remove atmospheric seeing and obtain diffraction limited observations \citep{Euclid10}.

For ground-based observatories, the varying refractive index in the turbulent layers of the atmosphere produces two effects.  It blurs the images of point sources and also introduces an instantaneous ellipticity at the 10\% level \citep{deVries}.  This can be viewed a superposition of speckles.  As the exposure time increases, these speckles de-correlate producing a circular distortion pattern.  Telescope optics and the camera distortion then induce an additional smoothing term and an anisotropic distortion across the camera field of view.  {The combined observed PSF size is roughly given by adding the size of the atmospheric PSF in quadrature with the size of the instrument PSF. Optimal instrumentation design therefore aims to minimize the instrument contribution so that the resulting PSF is atmosphere dominated}.  Image distortions are encompassed by the PSF as imaged by stellar point sources.    Using colour and size information, stars can be separated from galaxies and used to model the PSF.   For weak lensing measurements to date, the PSF model is typically an interpolation between stars, using a low order polynomial model to obtain a PSF at each galaxy location.  {This assumes however that the PSF does not vary rapidly on small scales \citep{PH08}.}

In this paper we measure and analyse the image distortion induced in short exposure observations by atmospheric aberrations.  This is of particular relevance to the PanSTARRS and upcoming LSST ground-based surveys.  Both these surveys will image more than 20,000 sq.\ degrees with LSST achieving a depth of $r \sim 27.5$ over 10 years.  As transient objects are a key science goal for both these ambitious surveys repeat short exposure observations are required.  For LSST each region of sky will be imaged close to 1000 times with 15 second exposures.  The current Pan-STARRS survey images each region of sky 5 times with 30 second exposures.  As both surveys progress a deep data set will be collected for optimal weak lensing analysis, but with such short exposure times the question is whether the rapidly varying atmospheric ellipticity will dominate the PSF degrading the ability to model the PSF in individual exposures using standard methods.

\cite{Wittman} first looked at this issue by analysing a set of 10 second and 30 second exposure images of a dense stellar field from the Subaru Telescope taken in one night.  The conclusion from this paper was that the spurious small scale power induced by the atmosphere would not be a significant factor in the co-added data of future massive surveys based on the finding that the atmospheric distortion in exposures separated by $\sim 120$ seconds was found to be uncorrelated.    This analysis was extended by \cite{Asztalos} with an analysis of 15 second exposure images from the Gemini South Telescope where evidence was presented for a correlation between the atmospheric distortion and the ground-based wind speed and direction with a consequence that the co-addition of exposures taken with the same observing site prevailing wind may not de-correlate as concluded by \cite{Wittman}.   

An investigation of the image ellipticity of atmospheric aberrations in simulated short exposure data is presented in \cite{deVries}.  Using ray-tracing and Fourier methods they simulate a turbulent atmosphere with a single layer Kolmogorov power spectrum, finding significant 10\% ellipticities in the instantaneous atmospheric PSF.  They also investigate the time dependence of the atmospherically induced ellipticities.  Assuming a constant wind speed they show the amplitude of the atmospheric ellipticity decreases with time $t$ as $t^{-1/2}$.  This simulation work has been extended by \cite{JeeTyson10} with simulated LSST short exposure images with a PSF that includes a six layer Kolmogorov power spectrum turbulence component and a ray-tracing optical distortion model.   They recover the \cite{deVries} result that as exposure time increases, the atmospheric PSF becomes rounder.  Using a Principal Component Analysis (PCA) they model the simulated short-exposure PSF across the LSST field-of-view and, keeping only the most significant 20 eigen-modes of that PCA analysis, they simulate what is described as a realistic PSF-convolved LSST image.  They conclude that a typical stellar density is sufficient to model the PCA derived PSF and demonstrate their ability to remove that PSF distortion from galaxy shape measurement.  By limiting the PSF to only 20 eigen-modes however, this analysis suppresses the random and high spatial frequency variation that we investigate in this study.

In this paper we assess the impact of atmospheric distortions on weak lensing surveys by analysing short exposure images of high density stellar regions with the wide-field MegaCam Imager on the Canada-France-Hawaii Telescope (CFHT).    It is of particular importance to revisit this issue as a significant improvement in the accuracy of weak lensing shape measurement has recently been found when data is analysed using individual exposures compared to the analysis of stacked exposures (Miller et al. in prep). Weak lensing data in the future will therefore not be analysed from a co-addition of exposures where all atmospheric turbulent effects have de-correlated, as it has been to date.  With a wide range of exposure time, wind speeds and directions we test the findings of \cite{deVries} that the atmospheric ellipticity decreases with time $t$ as $t^{-1/2}$ and also the findings of \cite{Asztalos} that the atmospheric ellipticity is correlated with the wind speed and direction.  Our results complement efforts to accurately simulate these effects (Chang et al. in prep) and inform the hardware and software design of future ground-based surveys accordingly.   This paper is organised as follows.  In Section~\ref{sec:data} we describe the data and analysis methods, presenting the results and analysis in Section~\ref{sec:res} and conclusions in Section~\ref{sec:conc}.

\section{Data and Methods}
\label{sec:data}
We analyse the CFHT PSF using a series of observations of dense stellar fields imaged and archived by the CFHT and listed in Table~\ref{tab:data}.  We select exposures taken in sub-arcsecond seeing conditions in the $i$-band over a range of exposure times.   Our main sample of 60 exposures imaged in 74 seconds were taken during three different observing runs sampling a wide range of wind speeds and directions.  We also analyse nine other samples with exposure times ranging from 1 second to 450 seconds with each set observed on different nights.

\begin{table}
\begin{center}
\begin{tabular}{l|l|l|l|l|}
Exptime & N Exp & Proposal ID  & Date\\ \hline
1.0     & 28    &   08AH22 & 2008-04-14/05-07\\ \hline
2.0     & 7     &   03BF99 & 2004-01-19/23 \\ 
        &       &   04BF28 & 2004-12-05/12 \\ 
        &       &   03BF09 & 2003-11-18/12-16 \\ \hline
7.5     & 3     &   08AQ98 & 2008-05-09/06-03\\ \hline
10.0    & 33    &   06BH48 & 2006-08-19/22\\ \hline
30.0    & 9     &   03BF99 & 2004-01-19/23\\ \hline
45.0    & 5     &   03BQ97 & 2003-09-30\\ \hline
74.0    & 60    &   08AH22 & 2008-04-14/05-08 \\ \hline
180.0   & 9     &   03BL06 & 2003-10-03 \\ \hline
300.0   & 11    &   08AH53 & 2008-02-13/14\\ \hline
450.0   & 12    &   07BC02 & 2007-11-14/01-10\\ \hline

\end{tabular}
\end{center}
\caption{Table of CFHT archive data used in this analysis listing the exposure time, total number of exposures used, the CFHT Proposal ID which can be used to query the CFHT Science Data Archive, and the range of observation dates.}
\label{tab:data}
\end{table}

The data were processed and stellar object catalogues produced as described in \cite{Hoekstra06} yielding a typical stellar density of 7 stars per square arcminute, {and in all datasets a significantly greater stellar density than available in extragalactic imaging survey data taken out of the galactic plane}.    We parameterize the PSF in terms of stellar ellipticity as measured by weighted quadrupole moments
\be
Q_{ij} = \frac{\int \, d^2\theta \, W(\bm{\theta}) \,
  I(\bm{\theta}) \, \theta_i
  \theta_j} {\int d^2\theta \, W(\bm{\theta}) \,I(\bm{\theta}) },
\label{eqn:quadmom}
\ee
where $I(\bm{\theta})$ is the surface brightness of the star, $\theta$ is the angular
distance from the star centre and $W$ is a Gaussian weight function of scale
length $r_g$, where $r_g$ is the measured dispersion of the PSF \citep{KSB}.  
For a
perfect ellipse, the weighted quadrupole moments are related to the
weighted ellipticity parameters $\varepsilon_\alpha$ by
\be
\left(
\begin{array}{c}
\varepsilon_1 \\
\varepsilon_2
\end{array}
\left)
= \frac{1}{R^2}
\right(
\begin{array}{c}
Q_{11} - Q_{22} \\
2Q_{12}
\end{array}
\right) \, ,
\label{eqn:ellipquad}
\ee
where $R$ is related to object size and given by
\be
R = \sqrt{Q_{11} + Q_{22}}\, ,
\ee
If the weight function $W(\bm{\theta}) = 1$ in equation~\ref{eqn:quadmom},
the ellipticity or polarisation  $|\varepsilon| = (1-\beta^2)/(1+\beta^2)$, where $\beta$
is the axial ratio of the ellipse \citep[see][]{Bible}.   We note that for the purposes of this study the quadrupole moment description of the PSF that we adopt is sufficient.  For a detailed weak lensing analysis however, many techniques now determine either a set of orthonormal 2D basis functions or a 2D pixel-based model to describe the PSF \citep{GREAT10}. Both these methods would be more challenging to model in the presence of high order spatial frequency variation.

In each exposure the variation of the PSF ellipticity $\varepsilon_i^{\rm PSF}$ and size $R$ across the field of view is considered to have two components; a smoothly varying second order polynomial over position, $\varepsilon_i^{\rm model}(x,y)$ and $R(x,y)$ for each chip and higher spatial varying residuals with 
\be
\delta \varepsilon_i = \varepsilon_i^{\rm PSF} - \varepsilon_i^{\rm model} \, ,
\ee
\be
\delta R^2 = R^2_{\rm PSF} - R^2_{\rm model}  \, .
\ee
With a typical number density of stellar objects imaged in an high-Galactic latitude field (40 per MegaCam chip, compared with over 700 in this analysis), the chip-wise second order polynomial model would be the most complex model that could be accurately fit to lensing survey data \citep{Rowe10}.

\section{Analysis and Results}
\label{sec:res}
\begin{center}
\begin{figure}
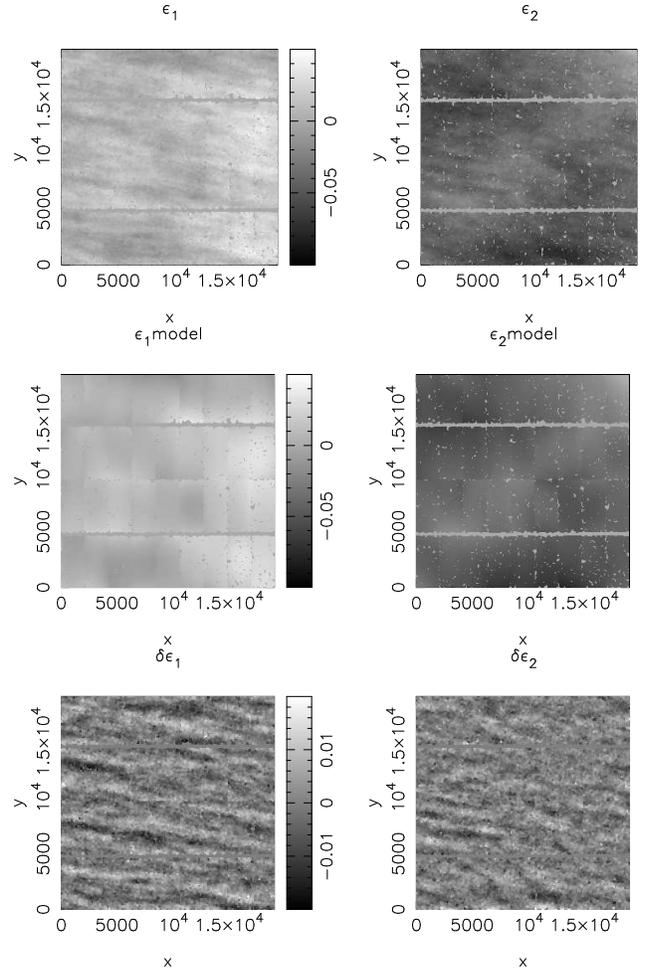

\epsfig{file=120data.ps,width=4.25cm,angle=270,clip=} 
\epsfig{file=120model.ps,width=4.25cm,angle=270,clip=}
\epsfig{file=120eres.ps,width=4.25cm,angle=270,clip=}
\caption{A typical PSF pattern for a 74 second exposure.  The left and right panels show the variation in the amplitude of the average PSF $\varepsilon_1$ and $\varepsilon_2$ respectively across the field of view.  The upper panels show the observed ellipticity variation, the middle panels show the second order polynomial model fit to the data.  The lower panels reveal the residual ellipticities $\delta \varepsilon_1$ and $\delta \varepsilon_2$ components.  The residuals show high spatial frequencies and a preferred direction.  The ellipticity amplitude in each panel is indicated by the greyscale shown in the vertical colour bar.}
\label{fig:psfex}
\end{figure}
\end{center}

\begin{center}
\begin{figure}
\epsfig{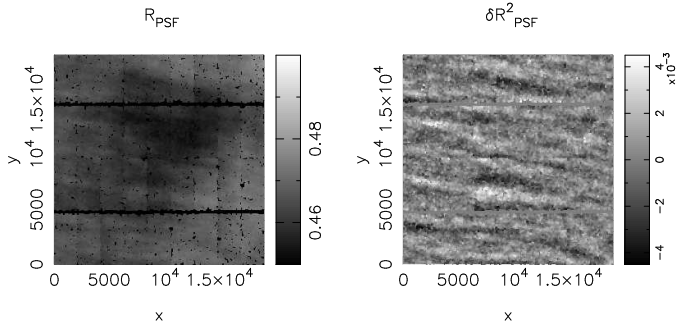} 
\caption{A typical PSF size pattern for the same 74 second exposure shown in Figure~\ref{fig:psfex}.  The left hand panel shows the variation in PSF size $R$ across the field with the greyscale in arcseconds. The right hand panel shows the residual variation in size after a second order polynomial model has been removed.  The ripple pattern in size follows the same structure as the ellipticity residuals.  The right hand greyscale is in units of $10^{-3}$ arcsec$^2$.
}
\label{fig:R2ex}
\end{figure}
\end{center}

Figure~\ref{fig:psfex} shows a typical PSF pattern for a 74 second exposure as imaged by MegaCam at CFHT.  The field of view is a square degree with a pixel scale of $0.186$ arcseconds and an $9 \times 4$ CCD pattern which is visible. The left and right panels show the variation in the amplitude of the average PSF $\varepsilon_1$ and $\varepsilon_2$ respectively across the field of view.
Each grid point in the greyscale map contains on average 5 stars and spans $0.3\times0.3$ arcmins.  In an image of a typical field, out of the Galactic plane, we would find $\sim 1$ useable star in every $\sim 5$ grid points ($\sim 0.4$ stars per sq arcmin).
The upper panels show the observed ellipticity variation.  The middle panels show the second order chip-wise polynomial model fit to the data.  The lower panels reveal the residual ellipticities $\delta \varepsilon_1$ and $\delta \varepsilon_2$ components.  Figure~\ref{fig:R2ex} shows the size variation for the same exposure with the size residuals (right hand panel) showing the same high spatial frequencies and a preferred direction as the ellipticity residuals.    For the rest of the paper we focus mainly on the ellipticity variation, but come back to the size variation in Section~\ref{sec:conc} where we show the high spatial frequency variation of PSF size is as detrimental to the weak lensing shape measurement as the variation in ellipticity.

\begin{center}
\begin{figure}
\begin{tabular}{cc}
\epsfig{file=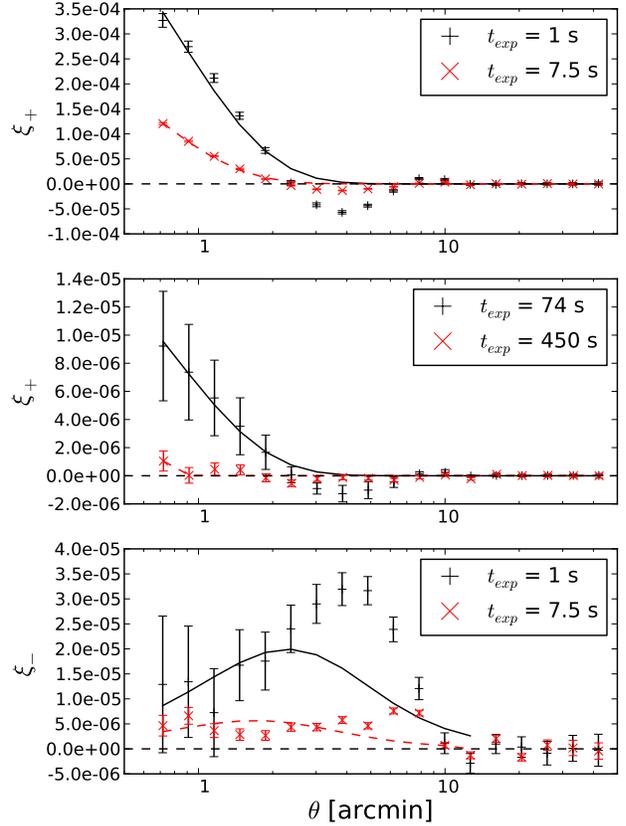,width=8.3cm,angle=0} &
\end{tabular}
\caption{The $\xi_+(\theta)$ and $\xi_-(\theta)$ correlation function estimates for the PSF ellipticity data, with best-fitting von K\'{a}rm\'{a}n models overlaid on the $\xi_+$ results {and the corresponding $\xi_-$ prediction plotted over the measured $\xi_-$ results} (see Section \ref{sect:model} {and Appendix \ref{app:vK}}).  Upper panel: the $t=1$s (solid line: best-fitting model) and $t=7.5$s (dashed line: best-fitting model) $\xi_+(\theta)$ results.  Middle panel: the $t=74$s (solid line: best-fitting model) and $t=450$s (dashed line: best-fitting model) $\xi_+(\theta)$ results.  Lower panel: the $t=1$s, and $t=7.5s$, $\xi_-(\theta)$ results (the longer exposure PSF patterns showed $\xi_-$ consistent with zero on all scales).}
\label{fig:corr}
\end{figure}
\end{center}

\vspace{-1.35cm}
In order to investigate the high spatial frequency variation of the PSF we measure the two-point correlation function of the residual PSF ellipticities in Figure~\ref{fig:corr}, showing the average systematic residual PSF correlation functions $\xi_+$ (upper two panels) and $\xi_-$ (lower panel) where
\be 
\xi_{\pm}^{\rm sys}(\theta) = \langle \varepsilon_t(\theta')  \varepsilon_t(\theta' + \theta) \rangle  \pm \langle \varepsilon_r(\theta')  \varepsilon_r(\theta' + \theta) \rangle
\ee
and $\varepsilon_{t,r}$ are the tangential and rotated ellipticity parameters rotated into the reference frame joining each pair of correlated objects.  The average is taken over all exposures in our sample split by exposure length.   
Figure~\ref{fig:corr} shows the amplitude of the residuals for four sets of decreasing exposure time revealing a characteristic shape to the correlation function that we find for all exposure times.  $\xi_+$ is positively correlated $\theta<2'$, anti-correlated $2'<\theta<7'$, then positively correlated until it becomes consistent with zero $\theta>15'$. This characteristic shape is also seen in the results of \cite{Wittman}.   $\xi_-$ is measured with significance only for the shortest exposures (shown lower panel) and is positively correlated for all scales $\theta < 15'$.  

The residual PSF correlation functions measured between consecutive $1s$, $10s$ and $74s$ exposures were found to be consistent with zero.  We can therefore conclude that the atmospheric turbulence distortion de-correlates in $<50$ seconds where the timescale is set by the CFHT MegaCam read-out, overheads, slew and acquisition time.  This is in agreement with \cite{Wittman} who set a de-correlation time $<120$ seconds, limited by the Subaru read time.

\vspace{-1.0cm}
\subsection{Comparison to a simple, isotropic turbulence model}
\label{sect:model}
The von K\'{a}rm\'{a}n model for isotropic atmospheric turbulence predicts a projected, two-dimensional power spectrum
\begin{equation}\label{eq:Pl}
P(l) \propto \left(l^2 + \frac{1}{\theta_{0}^2} \right)^{-11/6},
\end{equation}
where the angle $\theta_0$ defines the \emph{outer scale} of the spectrum (e.g.\ \citealp{sasiela94}). In this model, a physically-motivated modification of the scale-free Kolmogorov spectrum, turbulent structures in images of the sky become uncorrelated at separations greater than $\theta_{0}$.  As shown in Appendix \ref{app:vK}, the isotropic $\xi_+$ correlation function for such a spectrum may then be written as
\begin{equation}\label{eq:xipvK}
\xi_+(\theta) \propto \theta^{5/6} K_{-5/6}\left( 2 \pi \theta / \theta_0 \right),
\end{equation}
where $K_{\nu}(x)$ is the modified Bessel function of the second kind \citep{arfken05}.  Numerical approximation of these functions is simple via the useful series expansion of \citet{kostroun80}.  {The functional form of $\xi_-$ is given in equation \eqref{eq:xim}.  This expression is significantly more complicated than the $\xi_+$ case and was found to be prone to numerical instabilities.}

We now investigate whether the model given in equation \eqref{eq:xipvK} is able to reproduce the correlation function results seen in Figure~\ref{fig:corr}.  We restrict ourselves to fitting only the $\xi_+$ {model, which can be simply calculated and carries almost all the detectable signal. We go on to compare predictions for $\xi_-$ based on these fitting results to measurements of $\xi_-$ in the data}.
In Figure~\ref{fig:corr} we have overlaid the best-fitting von K\'{a}rm\'{a}n models to the $\xi_+$ correlation function estimates for each of the example $t=1$s, 7.5s, 74s \& 450s exposure data sets.  The Levenberg-Marquardt algorithm (e.g.\ \citealp{Numrec}) was used to determine the maximum-likelihood parameter fits to the overall amplitude of the expression in equation \eqref{eq:xipvK}, and for the outer scale length $\theta_0$. {In the lower panel of Figure~\ref{fig:corr} we use the best-fitting values of $\theta_0$ from the $\xi_+$ models to generate a von K\'{a}rm\'{a}n prediction for the $\xi_-$ signal in the t=1s and 7.5s exposure data.}

It can be seen that the von K\'{a}rm\'{a}n model gives a reasonable fit to the inner slope of $\xi_+$ for the three $t=1$s, 7.5s \& 74s exposure data sets in Figure \ref{fig:corr}. For these three sets we also found broadly consistent best-fitting values of the outer scale in the range $\theta_0 = 2.62$--$3.22'$.  These results suggest a turbulent origin for the small-scale correlated ellipticity measurements in these image, but in all three cases the von K\'{a}rm\'{a}n model fails to capture both the trough on scales $\theta \simeq 4'$, and the second peak in correlation near $10'$.
{The discrepancy is also marked in the comparison between the $\xi_-$ model and the data.  Here a very distinct excess `bump' is seen, again peaking around $\theta \simeq 4'$ in the 1s exposure time data.  However, as the amplitude of the $\xi_-$ correlation is consistently smaller than that of $\xi_+$, these results are noisier.}
  {We discuss potential origins for these features in Section~\ref{sec:conc}.}  For the $t=450$s data we found a model amplitude consistent with zero, indicative of the weakness of the signal for this long exposure data.


\subsection{Dependence of the amplitude of the atmospheric distortion on exposure time}
\label{sec:timedep}
\begin{center}
\begin{figure}
\epsfig{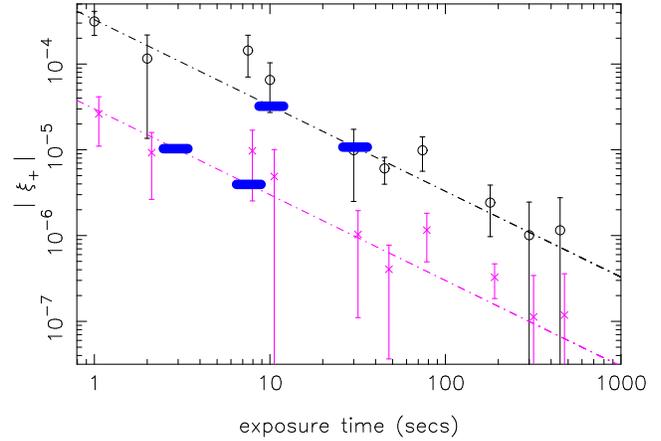} 
\caption{The evolution with exposure time of the atmospheric turbulence as measured by the two-point correlation function $\xi_+$ at two different angular scales; 0.7 arcmin(black upper) and 4.9 arcmin (red lower). 
The best-fit de Vries et al.(2007) prediction is shown dot-dashed for each angular scale.  This can be compared to {the two pairs of thick horizontal bars over plot for each angular scale.   These bars indicate the}
expected CFHTLenS cosmological signal for two tomographic bins with a mean redshift $z = 0.37$ and $z=0.54$. determining the exposure time below which the cosmological signal becomes lower than the residual atmospheric PSF signal.}
\label{fig:time}
\end{figure}
\end{center}

A key result from the simulations of \cite{deVries} showed that the ellipticity introduced by atmospheric aberrations decreases with time as $t^{-1/2}$ assuming a constant wind speed.   The two-point correlation function of the atmospheric variations $\xi_+$ is therefore expected to decrease in amplitude as $t^{-1}$.  Figure~\ref{fig:time} shows the average two-point correlation function $|\xi_+|$ as a function of the exposure time.  Results are shown for two representative angular scales with $\xi_+$ measured at $\theta = 0.7'$ (upper) where the von K\'{a}rm\'{a}n atmospheric turbulence model provides a good fit to the data, and at $\theta = 4.9'$ which corresponds to the first negative dip in the correlation function.  We find that the \cite{deVries} model (shown dot-dashed) is a good fit to the data but note that for this analysis, we were unable to select a significant sized sample of exposures with a constant wind speed as is assumed in the \cite{deVries} theory.  We conclude that the scatter we see in the results must be in part driven by the range of wind speeds within our sample.  

Figure~\ref{fig:time} also allows us to compare the amplitude of the atmospheric turbulence to the expected cosmological signal in the CFHT Lensing Survey (CFHTLenS\footnote{CFHTLenS analyses data from the CFHT Legacy Survey, using both the Wide, Deep and Pre-imaging surveys}).  For both angular scales $\theta$ shown we over-plot two horizontal bars that indicate the amplitude of the two-point shear correlation function $\xi_+(\theta)$ for two low redshift tomographic bins with a mean redshift of $z = 0.37$ and $z= 0.54$ assuming a WMAP7 cosmology \citep{WMAP7}.  We should note here that one cannot directly relate the PSF ellipticity correlation with that expected from shear, as the effect of the PSF on the shear measurement depends on the relative size of galaxy such that it would negligible for large galaxies or potentially amplified for galaxies comparable to or smaller than the PSF \citep{PH08}.  This comparison does however demonstrate that even for the lowest redshift tomographic bin, the residual systematic error from atmospheric turbulence is well below the cosmological signal for the 600 second CFHTLenS exposures.
\begin{center}
\begin{figure}
\begin{tabular}{cc}
\epsfig{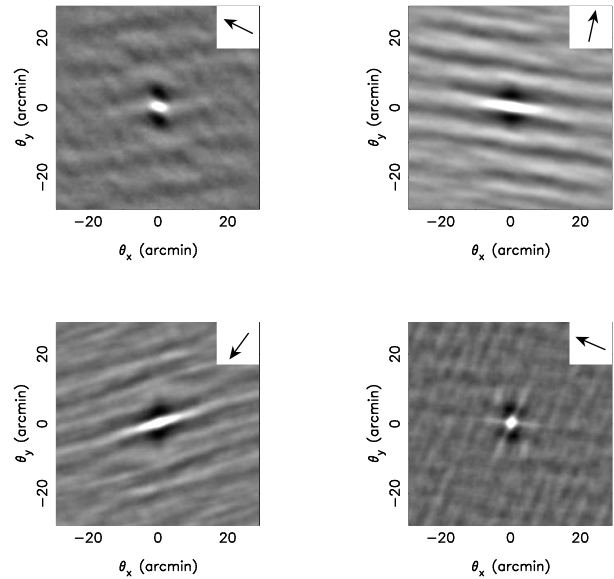} &
\end{tabular}
\caption{The $\xi_+$ residual correlation as a function of 2D \emph{vector} separation $\bm{\theta}$ {for four example 74 second exposures.  The dominant direction of the PSF residuals can be compared to the wind direction shown as an arrow inset.  The upper left panel shows an example exposure with a wind speed of $14 \textrm{ms}^{-1}$. The upper right panel shows the exposure also shown in Figure~\ref{fig:psfex} and has a wind speed of $6 \textrm{ms}^{-1}$.  The lower left panel has a wind speed of $2 \textrm{ms}^{-1}$ and the lower right panel has a wind speed of $11 \textrm{ms}^{-1}$.}}  
\label{fig:2Dcorr}
\end{figure}
\end{center}

\vspace{-1.5cm}
\subsection{Dependence on wind speed and direction}\label{sect:windspeed}
In order to compare the turbulence patterns observed to the wind direction we quantify the direction of the atmospheric turbulence using correlation as a function of vector separation and direction on the sky, rather than the commonly-used, azimuthally-averaged functions $\xi_+(\theta) = \xi_+(|{\bm{\theta}}|)$.  We calculate the $\xi_+$ residual correlation as a function of vector angular separation ${\bm{\theta}}$ using the Discrete Fourier Transform to find the power spectrum of the $\delta \varepsilon_1$ and $\delta \varepsilon_2$ images (examples of which are shown in Figure~\ref{fig:psfex}), then employing the Wiener-Khinchin theorem to generate images of $\xi_+({\bm{\theta}})$.  

{Four example results are shown in Figure~\ref{fig:2Dcorr} revealing a non-isotropic function. The example exposure shown in Figure~\ref{fig:psfex} is the same exposure presented in the upper right hand panel of Figure~\ref{fig:2Dcorr}.  A comparison of these two Figures shows that the elliptical $\xi_+(|{\bm{\theta}}|)$ has a central dipole with a major axis along the direction of the ripples seen in the $\delta \varepsilon_i$ maps of Figure~\ref{fig:psfex}. For other exposures however we see what appears to be a superposition of ripple patterns (see for example the upper left and lower right panels of Figure~\ref{fig:2Dcorr}), potentially arising from different turbulent layers in the atmosphere.  In these cases the orientation of the central dipole indicates the average ripple direction.  Based on these comparisons over the full data set we use the orientation of the central dipole to determine an effective ripple orientation and we measure this orientation using the quadrupole moments in equation~\ref{eqn:ellipquad}) with $r_g = 2.3$ arcmin.  
Applying this method to all exposures with $t \le 74$s, where the amplitude of the atmospheric turbulence is sufficiently high, we can then compare this ripple direction to the wind direction relative to the image (shown by the arrow in the upper inset in Figure~\ref{fig:2Dcorr}), calculated as described in Appendix \ref{app:wind}.}
\begin{center}
\begin{figure}
\epsfig{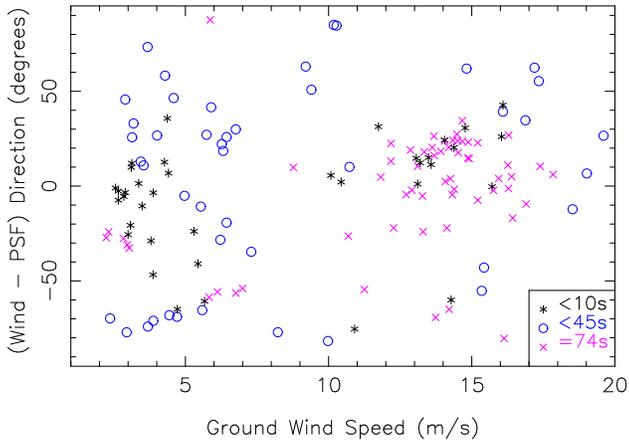} 
\caption{A comparison of wind direction and PSF residual direction as a function of the ground wind speed measured at the start of each observation for three samples grouped by exposure time with $t \le 10$s (stars), $10$s$ > t \le 45$s (circles) and $t=74$s (crosses).}
\label{fig:Wind_Dir_Speed}
\end{figure}
\end{center}

\vspace{-0.75cm}
Figure~\ref{fig:Wind_Dir_Speed} shows a compilation of results for three sets of data; $t \le 10$s (stars), $10$s$ > t \le 45$s (circles) and $t=74$s (crosses) comparing wind direction and PSF or ripple orientation.  We find that in general there is little evidence of strong correlation between wind direction and PSF orientation for the range of exposure times tested.
This result is in disagreement with \cite{Asztalos} who found a relationship between wind direction and PSF ellipticities for wind speeds ranging from 2-6 m/s over 4 consecutive nights.  We note that we could have drawn a similar conclusion had we analysed a smaller set of exposures, as the points that cluster in Figure~\ref{fig:Wind_Dir_Speed} are typically taken on the same night.  This clustering with observation date, if sets of observations were to be analysed in isolation, might well suggest a preferred wind-PSF orientation which we do not find when we analyse a large set of data spanning many years of CFHT imaging.
\begin{center}
\begin{figure}
\epsfig{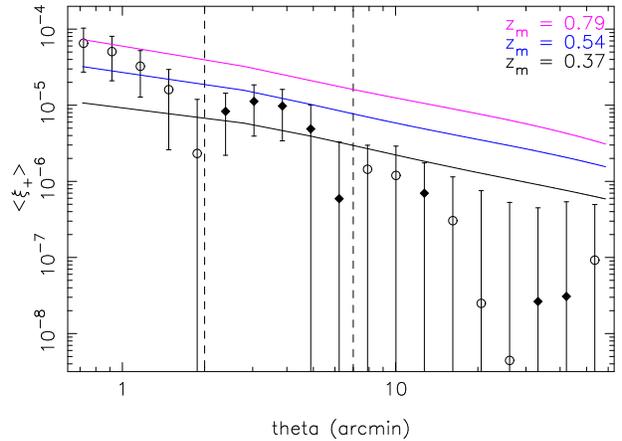} 
\caption{The measured residual ellipticity correlation in our $t=10$s data set compared to the expected WMAP7 cosmological two-point shear correlation signal for three tomographic bins with a mean redshift $z = 0.37$, $z=0.54$ and $z= 0.79$ (solid lines, the lowest amplitude corresponds to the lowest redshift bin).  The data oscillates between positive and negative correlation which we indicate on this log-log plot using filled
points where the data is negative which is primarily between the two dashed vertical lines.}
\label{fig:cosshearcomp}
\end{figure}
\end{center}
\vspace{-0.75cm}

\section{Conclusions:  Impact for weak lensing observations}
\label{sec:conc}
In this study we have analysed short exposure observations of dense stellar fields to quantify the high spatial frequency variation of the PSF which we attribute to atmospheric effects.  We have shown that in short exposures with $t\le 30$s, the atmosphere contributes a significant correlated anisotropy to the PSF on angular scales $\theta < 10'$ that would dominate the cosmological signal measured in the lower redshift bins of a cosmological tomographic analysis.  On these angular scales the high spatial frequency of the atmospheric aberration is too rapid to model with a typical stellar density and standard methods.  

{ This does not mean, however, that multiple short exposure images cannot not be combined in a way that reduces the cumulative impact of this (random) atmospheric anisotropy on estimates of gravitational shear in the final analysis.   It was found in this study that the turbulent patterns in consecutive exposures  are statistically uncorrelated when separated by timescales $\le 50$s.   Our observations motivate the combination of information from multiple short exposure images in a way that takes advantage of this fact.  Examples of such approaches might be cross-correlating shear estimates from different exposures to estimate the shear autocorrelation, or combining images into a stacked, co-added image.  If stacking, great care must taken to retain control and knowledge of changes in the PSF due to interpolation and stacking of dithered data. \citealp*{roweetal11imcom} present an example of a linear image stacking approach that seeks to preserve this information.}

{ As well as finding that consecutive exposures are uncorrelated, we have confirmed the predictions of \cite{deVries} in finding that the amplitude of atmospheric distortion patterns decreases with time as $t^{-1/2}$}.  Our results also show that this effect is not a significant source of error for the CFHTLenS survey where for the first time a lensing survey is analysing the 600 second exposures individually (Miller et al in prep), as compared to previous analyses using a stack \citep{Fu08}.

Figure~\ref{fig:cosshearcomp} compares the measured residual ellipticity correlation in our $t=10$s data set with the expected cosmological signal for three tomographic bins with a mean redshift $z = 0.37$, $z=0.54$, $z=0.79$.  This is the closest data set analysed to the proposed $t=15$s exposures for LSST.
This figure shows that our findings agree to some extent with \cite{JeeTyson10}, who focus on angular scales $\theta > 10'$ and find that the PSF can be modeled to high accuracy on these scales.  For smaller angular scales, however atmospheric turbulence effects will be a significant source of systematic error in a weak lensing analysis if not corrected for.  Furthermore, as a result of the anisotropic nature of the residual correlation function (shown in Figure~\ref{fig:2Dcorr}) the effect of atmospheric turbulence may leak to larger scales.  

For a future generation all-sky lensing survey \citet{AmaraRef} derive a requirement on the variance of systematic errors to be below $\sigma^2_{\rm sys} < 10^{-7}$ such that experiments are limited by statistical noise rather than systematic errors (see also \citet{VW06} for a similar conclusion).  To reach this goal \cite{PH08} show that this requires $\sigma[\varepsilon_{\rm PSF}] \lesssim 10^{-3}$ for each ellipticity component and $\sigma[R^2_{\rm PSF}] \lesssim 10^{-3} R^2_{\rm PSF}$ {for $\varepsilon_{\rm PSF} \simeq 0.05$ and a typical galaxy/PSF size ratio of $R_{\rm gal}/R_{\rm PSF} \gtrsim 1.5$}.  For the $t=10$s data we measure  
\be
\sigma[\varepsilon_{\rm PSF}]= \sqrt{\langle \delta \varepsilon _i^2 \rangle} = (9.9 \pm 1.7) \times 10^{-3} \, \label{eq:sigepsf}
\ee
\be
\frac{\sigma[R^2_{\rm PSF}]}{R^2} = \frac{\sqrt{\langle \delta R^2 \rangle}}{\langle R^2 \rangle} = (9.6 \pm 2.0) \times 10^{-3} \,  \label{eq:sigR2psf}
\ee
where the error is calculated from the scatter between the 33 exposures in the $t=10$s data set showing the variance within the set, not the error on the mean.  These measurements are an order of magnitude larger in both size and ellipticity compared to the requirements quoted in \citet{AmaraRef}.

Using equation (15) in \cite{PH08} we can calculate the variance of the systematic errors we would expect from the residual atmospheric size and ellipticity variation that we measure from the $t=10$s data set: { we find $\sigma^2_{\rm sys} \sim 10^{-5}$.  This value can be interpreted as the systematic variance produced in multiple 10s exposures if: (i) the turbulent PSF cannot be modelled due to insufficient stellar density; and (ii) \emph{no attempt} is made to combine shear estimates from multiple exposures for which the turbulence patterns are mutually statistically independent.  We note that this calculation depends on the accuracy of both PSF ellipticity and PSF size characterisation, and both these are affected by atmospheric turbulence}.

{We note that while the expectation of the turbulent contribution to $\varepsilon_{\rm PSF}$ across multiple exposures is zero, assuming any static component is removed, this is \emph{not} the case for $R^2_{\rm PSF}$.  This has implications for plans to cross-correlate using only shear estimates from different exposures to suppress PSF model errors.   The utility of such a scheme is that the ellipticity cross  terms $\langle \delta \varepsilon_i \delta \varepsilon_j \rangle$ (for $i \ne j $) have zero expectation value, and the variance of this product is the square of the individual variances.  This product variance decreases  $\propto 1/N^2_{\textrm{exp}}$, where $N_{\textrm{exp}}$ is the total number of exposures being cross-correlated, and therefore $\sigma [\varepsilon_{\rm PSF} ]$ decreases rapidly as $\propto 1/N_{\textrm{exp}}$ in the cross-correlation. }

{However, the same approach does not reduce uncertainty in $R^2_{\rm PSF}$ at the same rate because this quantity has an
unknown expectation value which must be characterized.  Instead, $\sigma [R^2_{\rm PSF}] \propto 1 / \sqrt{N_{\textrm{exp}}}$.   This differing behaviour in the combination of multiple exposures must be considered when comparing the results in equations \eqref{eq:sigepsf} \& \eqref{eq:sigR2psf} to the required levels for the estimation of cosmic shear.  The characterization of turbulent variation in $R^2_{\rm PSF}$ therefore assumes added importance for upcoming surveys.  Combined with $N_{\textrm{exp}}$, the single-exposure value quoted in equation \eqref{eq:sigR2psf} can be used to estimate this important effect for surveys using multiple exposures of duration $t \simeq 10$s, assuming similar atmospheric conditions to those at CFHT}.


Both the results in this paper and \citet{Wittman} show an oscillatory pattern in the residual $\xi_+$ correlation function at $\theta > 2.5$ arcminutes, a pattern which is not consistent with the model predicted by a simple, isotropic von K\'{a}rm\'{a}n spectrum (see Figure \ref{fig:corr}).   \citet{Wittman} describes this ringing as an artifact of the interpolation scheme used to model the PSF on each individual chip which for the Suprime-Cam camera used in these observations has a chip size of $13.65' \times 6.83'$.   In this analysis we apply a similar, low-order chipwise PSF model, where for MegaCam the chip size is $12.70' \times 6.35'$. We thus agree with \citet{Wittman} that one possible origin for the `ringing' is over-subtraction of a chipwise PSF model for the telescope optics, as the first dip occurs at a close to half the { shorter dimension of the chip}.  

However, it is less clear why chipwise fitting would cause an excess at $10'$.  In addition, looking at the $\xi_+({\bm{\theta}})$ (i.e.\ not circularly averaged) two-point correlation functions (e.g.\ Figure~\ref{fig:2Dcorr}) we see that the oscillatory pattern is anisotropic in nature but often not aligned with the $x$-$y$ chip grid.  This alignment would be something to be expected if it were an artifact of the chipwise PSF model removal.    We therefore offer an alternative suggestion that these ripples have an atmospheric origin, related to anisotropy in the atmospheric power spectrum (c.f.\ Figure \ref{fig:2Dcorr}), which cannot be generated by simple isotropic models such as the von K\'{a}rm\'{a}n. { This conclusion is supported by the finding that the residual patterns are uncorrelated between exposures.}  However, without further data, or access to imaging with chips of differing dimensions, it remains difficult to say definitively which interpretation is correct.

Although not performed explicitly in this study, it is worthwhile mentioning the separation of the correlated turbulence signal into `E' and `B' mode contributions (see, e.g., \citealp{CNPT02}).   The atmospheric PSF pattern for a turbulent sky may contain both E- and B-modes; if this PSF is not properly modelled any shear catalogue will then contain an imprint of this pattern.  An E/B-mode decomposition of the cosmic shear signal on the scales where atmospheric turbulence patterns are correlated might therefore detect this effect in the form of a non-zero B-mode.   However, telescope optical abberrations, which also contribute strongly to PSF anisotropy across most angular scales of interest for lensing, tend to generate a more significant E-mode than B-mode \citep{Hoekstra03}.  In addition, cosmological effects can also generate B-modes (e.g., \citealp{SchvWM02,valeetal04}).  The absence of strong B-modes in cosmic shear measurements therefore remains a problematic indicator of the absence of systematic errors due to PSF anisotropy, and the potential presence of B-modes in turbulent PSF patterns does not alter this fact.

Finally, when considering our full sample of data, we have found little evidence for a strong and consistent relationship between the preferred direction of anisotropic correlations in atmospheric distortion and the ground-based wind velocity.  We do not necessarily expect to find such a relationship, as the all the turbulent layers responsible for the atmospheric PSF pattern are not necessarily correlated with the ground conditions.
However, we do find that consecutive observations seem to yield similar angular offsets between wind direction and turbulent PSF pattern orientation: such effects might lead to an over-interpretation of the correlation between wind direction and PSF pattern orientation in more limited data sets.  In all, these results do not strongly motivate the use of the ground-based wind direction and speed as an input to a PCA-type model of the atmospheric part of the PSF on short exposure data. However, it may still be useful in modelling out repeatable PSF distortions due to telescope motion under wind (see, e.g., \citealp{JarvisPSF}).   { Once again, this motivates data analysis strategies that take advantage of the turbulent pattern being uncorrelated between successive exposures}.

\section{Acknowledgements}
We thank Chihway Chang, Phil Marshall and the rest of the CFHTLenS collaboration for convincing us to write this analysis into a paper and for many useful discussions along the way.  We also thank { the anonymous referee for helpful comments} and the CFHT QSO team and PIs of the various data sets used in this analysis; Jerome Bouvier, Catherine Dougados, Eugene Magnier, Alan McConnachie and John Johnson.
CH \& BR acknowledge support from the European Research Council under the EC FP7 grant numbers 240185 (CH) \& 240672 (BR).  HH acknowledges support from Marie Curie IRG grant 230924 and the Netherlands Organisation for Scientific Research grant number 639.042.814.  TE is supported by the Deutsche Forschungsgemeinschaft through project ER 327/3-1 and the Transregional Collaborative Research Centre TR 33 - "The Dark Universe.  TDK was supported by a RAS 2010 Fellowship.
This study uses data from the CFHT Science Data Archive hosted and supported by the Canada Astronomy Data Centre operated by the National Research Council of Canada with the support of the Canadian Space Agency.

{\small Author contributions: All authors assisted with the development and writing of this paper.  CH and BR co-led the statistical analysis, HH conceived the project and led the data mining, catalogue production and data analysis, LM motivated the project and led the visualization of the effect.}

\appendix

\section{Correlation functions for the Von K\'{a}rm\'{a}n power spectrum}\label{app:vK}
As in the case of cosmological shear, the ellipticity correlation functions $\xi_+(\theta)$ and $\xi_-(\theta)$ \citep[see][]{Bible} due to atmospheric effects can be related to power spectrum of the von K\'{a}rm\'{a}n model as
\begin{eqnarray}
\xi_+ (\theta) &\propto & \int_0^{\infty} l \textrm{d} l  J_0(2 \pi l \theta )\left[l^2 + \frac{1}{\theta_0^2} \right]^{-11/6} \\
\xi_- (\theta) &\propto & \int_0^{\infty} l \textrm{d} l J_4(2 \pi l \theta) \left[l^2 + \frac{1}{\theta_0^2} \right]^{-11/6}  ,
\end{eqnarray}
where $J_{\nu}(x)$ is the $\nu$-th order Bessel function, and where we are assuming that the spatial power spectrum of ellipticity distortions due to turbulence follows $P(l)$ as given in equation \eqref{eq:Pl}.  Using these expressions, we calculate the $\xi_+$ correlation function as 
\begin{equation}\label{eq:xip}
\xi_+(\theta) \propto \theta^{5/6} K_{-5/6}\left( 2 \pi \theta / \theta_0 \right),
\end{equation}
where $K_{\nu}(x)$ is the modified Bessel function of the second kind \citep{arfken05}.
The expression for $\xi_-$ is more complicated:
\begin{eqnarray}
\xi_-(\theta) & \propto &  
\theta^{-7/6} \left( 6 \theta_0^2 + \pi^2 \theta^2 \right) I_{-17/6}(2 \pi \theta/\theta_0)  \nonumber \\
& - &  \frac{13 \pi }{6} \theta_0 \theta^{-1/6} I_{-11/6} (2 \pi \theta / \theta_0) \nonumber \\ 
& + &  \theta^{-7/6} \left(\frac{5}{36}\theta^2_0 - \pi^2 \theta^2 \right) I_{17/6}(2 \pi \theta / \theta_0)  \nonumber \\
& + &  \frac{13 \pi}{6} \theta_0 \theta^{-1/6} I_{23/6}(2 \pi \theta / \theta_0) \nonumber \\
 & - & \theta_0^{17/6}\theta^{-4}\left(\frac{385}{432} \theta_o^2 - 70 \pi^2 \theta^2 \right) \Gamma(-7/6) \label{eq:xim},
\end{eqnarray}
where $I_{\nu}(x)$ is the modified Bessel function of the first kind and $\Gamma(x)$ is the Gamma function \citep{arfken05}.  The functions $I_{\nu} (x)$ diverge rapidly for $x > 1$, and the expression in equation \eqref{eq:xim} shows signs of numerical instability for $\theta \gtrsim 4 \theta_0$.  In this study we therefore make a crude approximation to avoid this behaviour.  We manually set $\xi_-(\theta) =0$ where $\theta > 4.2 \theta_0$ for the von K\'{a}rm\'{a}n power spectrum model when investigating or plotting $\xi_-$.

\section{The projected direction of ground wind in images on the celestial sphere}\label{app:wind}
We consider two general points with altitude-azimuth coordinates $(a_1, A_1)$ and $(a_2, A_2)$ in the horizontal coordinate system, which uses the observer's local horizon as the fundamental plane.  As for any two points on the coordinate sphere, a single great circle may be constructed that passes through $(a_1, A_1)$ and $(a_2, A_2)$.  We label as $\sigma$ the angular size of the arc between these two points on the great circle. 
Using the spherical law of cosines, 
it is straightforward to show that $\sigma$ is given by
\begin{eqnarray}\label{eq:Appsig}
\cos{\left(\sigma\right)}& =& \cos{\left(a_1\right)} \cos{\left(a_2\right)} \cos{\left(A_1 - A_2\right)} \nonumber \\
& +& \sin{\left(A_1\right)} \sin{\left(A_2\right)}.
\end{eqnarray}
This result will be used below to derive the projected direction of ground wind on telescope images.

The pixel $y$-axis of the MegaCam instrument mounted on CFHT rotates to align with the celestial North Pole.  In the horizontal coordinate system defined at the telescope, celestial North has the altitude-azimuth coordinates $\textbf{N} = (\phi, 0) $ where $\phi$ is the geographic latitude of Mauna Kea.  We define the telescope pointing direction (i.e.\ the direction of the image field centre) as $\textbf{T} = (a_{\textrm{t}}, A_{\textrm{t}})$ in this coordinate system.   The points $\textbf{N}$ and $\textbf{T}$ on the celestial sphere can be connected by the arc of a great circle of constant Right Ascension.  The $y$-axis of CFHT images will align with this Meridian. 

The direction of ground wind is measured at CFHT and supplied with the
astronomical images in degrees relative to compass North along the ground.  In the horizontal coordinate system at CFHT this direction can therefore be expressed as $\textbf{W} = (0, A_{\textrm{w}})$, where $A_{\textrm{w}}$ is the measured wind direction.   We may then define a second great circle passing through $\textbf{W}$ and $\textbf{T}$.  The angle $\theta$ at which this great circle intersects the Meridian through $\textbf{T}$ and $\textbf{N}$ is therefore the projected direction of ground wind on the telescope image, relative to the CCD $y$-axis.

The spherical law of cosines gives the following result for this projected wind direction angle $\theta$:
\begin{equation}\label{eq:Apptheta}
\cos{\left(\theta\right)} = \frac{ \cos{\left( \gamma \right)} - \cos{\left( \alpha \right)} \cos{\left( \beta \right)}}{\sin{\left( \alpha \right)} \sin{\left( \beta \right)} },
\end{equation}
where from equation \eqref{eq:Appsig} we have
\begin{eqnarray}
\cos{\left( \alpha \right)} & = & \cos{\left( a_{\textrm{t}} \right)} \cos{\left( A_{\textrm{w}} - A_{\textrm{t}} \right)}, \\
\cos{\left( \beta \right)} & = & \cos{\left( a_{\textrm{t}} \right)} \cos{\left(\phi \right)} \cos{\left(A_{\textrm{t}}\right)} + \sin{\left( a_{\textrm{t}} \right)}\sin{\left( \phi \right)}, \\
cos{\left( \gamma \right)} & = & \cos{\left( A_{\textrm{w}} \right)} \cos{\left( \phi \right)}.
\end{eqnarray}
However, the expressions above do not provide the \emph{sign} of $\theta$, i.e.\ whether $\textbf{W}$ lies to the West or East of the Meridian through $\textbf{T}$ and $\textbf{N}$, the $y$-axis on the telescope image.  This information is needed to correctly calculate the angular separation between the wind direction and primary PSF residual direction as described in Section \ref{sect:windspeed}.

The positive direction on the CCD $x$-axis of MegaCam can be associated with positive \emph{hour angle} $H$ in the equatorial coordinate system.  If we define the positive $\theta$ direction as that moving clockwise from the $y$-axis, then it is straightforwardly seen that
\begin{equation}\label{eq:Appsigntheta}
\textrm{sgn}{\left(\theta \right)} = \textrm{sgn}{\left( H_{\textrm{w}} - H_{\textrm{t}} \right)}
\end{equation}
where $\textrm{sgn}(x)$ is the signum function, $H_{\textrm{w}}$ is the hour angle of the wind direction vector $\textbf{W}$ and $H_{\textrm{t}}$ is the hour angle of the telescope pointing direction $\textbf{T}$ at the local sidereal time of observation.  The telescope hour angle $H_{\textrm{t}}$ is provided by CFHT as metadata alongside all astronomical images, so it remains solely to determine $H_{\textrm{w}}$.

Using standard formulae for the conversion of horizontal to equatorial spherical coordinates, the declination $\delta_{\textrm{w}}$ of the wind vector $\textbf{W}$ is given by
\begin{equation}
\sin{\left( \delta_{\textrm{w}} \right)} = \cos{\left( \phi \right)} \cos{\left( A_{\textrm{w}} \right)}.
\end{equation}
The hour angle of $\textbf{W}$ is then given by
\begin{eqnarray}
H_{\textrm{w}} & = & \arcsin{\left[ - \frac{\sin{\left( A_{\textrm{w}} \right)}}{\cos{\left( \delta_{\textrm{w}} \right)}} \right]} \nonumber \\
 & = & \arcsin{\left[ - \frac{\sin{\left( A_{\textrm{w}} \right)}}
{\sqrt{1 - \cos^2{\left( \phi \right)} \cos^2{\left( A_{\textrm{w}} \right)} }} \right]}.
\end{eqnarray}
When substituted into equation \eqref{eq:Appsigntheta} this allows the sign of $\theta$ to be calculated.  In combination with the determination of the magnitude of $\theta$ from equation \eqref{eq:Apptheta}, the projected direction of ground wind on a telescope image pointing in the direction $\textbf{T}$ is fully specified.

\bibliographystyle{mn2e}
\bibliography{ceh_2010}
\label{lastpage}

\end{document}